# DRAFT- FLOW BOILING OF WATER ON NANOCOATED SURFACES IN A MICROCHANNEL

Hai Trieu PHAN [a,b], Nadia CANEY [a], Philippe MARTY [a], Stéphane COLASSON [b], Jérôme GAVILLET [b]

[a] LEGI, BP 53, 38041 Grenoble Cedex 9, FRANCE
[b] LITEN, CEA Grenoble, 17 rue des martyrs, 38054 Grenoble Cedex 9, FRANCE
Email: hai-trieu.phan@cea.fr

## ABSTRACT

Experiments were performed to study the effects of surface wettability on flow boiling of water at atmospheric pressure. The test channel is a single rectangular channel 0.5 mm high, 5 mm wide and 180 mm long. The mass flux was set at 100 kg/m² s and the base heat flux varied from 30 to 80 kW/m². Water enters the test channel under subcooled conditions. The samples are silicone oxide (SiOx), titanium (Ti), diamond-like carbon (DLC) and carbon-doped silicon oxide (SiOC) surfaces with static contact angles of 26°, 49°, 63° and 103°, respectively. The results show significant impacts of surface wettability on heat transfer coefficient.

## NOMENCLATURE

- $A_h$    heat transfer area, m²
- $C_p$    specific heat (J/kg K)
- $D_h$    hydraulic diameter, m
- $g$    gravity (m²/s)
- $G$    mass flux (kg/m² s)
- $h$    heat transfer coefficient, W/m² K
- $h_{lv}$    latent heat of vaporisation, J/kg
- $I$    current, A
- $\dot{m}$    mass flow, kg/s
- $q$    heat flux, W/m²
- $q_{loss}$    lost heat flux, W/m²
- $N$    number of sections, -
- $T$    temperature, °C or K
- $\overline{T}$    average temperature, °C or K
- $V$    voltage, V
- $x$    vapour quality, -
- $\Delta x$    variation of vapour quality, -
- $z$    z coordinate (m)

**Greek symbols**
- $\lambda$    thermal conductivity (W/m K)
- $\mu$    dynamic viscosity (Pa s)
- $\rho$    density (kg/m³)
- $\sigma$    surface tension (N/m)
- $\theta$    static contact angle (°)
- $\theta_a$    advancing contact angle (°)
- $\theta_r$    receding contact angle (°)
- $\Delta\theta$    contact angle hysteresis(°)

**Subscripts**
- f    fluid
- i    position on the sample surface
- in    at the inlet of the test section
- l    liquid
- out    at the outlet of the test section
- s    saturation
- v    vapour
- w    wall

**Dimensionless numbers**
- Co    confinement number (Eq (1))
- Nu    Nusselt number $hD_h/\lambda$
- Pr    nombre de Prandtl ($\mu C_p/\lambda$)
- Re    Reynolds number $GD_h/\mu_l$
- $z^+$    reduced length $z/D_h \mathrm{Re}\,\mathrm{Pr}$

## 1. INTRODUCTION

Microchannels and minichannels are channels with small hydraulic diameters which are especially desirable for size reduction of heat exchange devices in microelectronic industry for example. In order to distinguish between macro and microscale flow boiling, the threshold to confined bubble flow is one of the most widely used criterions. Following the classification by Kew and Cornwell [1], channels are classified as microchannels if $Co \geq 0.5$, where $Co$ is the confinement number defined as:

$$Co = \left[\frac{\sigma}{g(\rho_l - \rho_v)D_h^2}\right]^{1/2} \qquad (1)$$

Kandlikar and Grande [2] suggested another classification based on the hydrodynamic diameter: conventional channels ($D_h > 3$ mm), minichannels (200 μm $< D_h <$ 3 mm) and microchannels ($D_h <$ 200 μm).

 

In microchannel flow boiling, the heat transfer coefficient can significantly decrease at low values of vapour quality due to intermittent dryout, which refers to an unstable breakdown of the liquid film in contact with the wall [3-5]. Intensification of the heat transfer coefficient and delay of the intermittent dryout are two main objectives of research in recent years [6].

At small scales, capillary effects become important and may play a significant role in the heat transfer mechanism. For instance, Phan *et al.* [7] showed that surface wettability affects the pool boiling heat transfer significantly. Modifying the surface wetting is thereby a reasonable solution to achieve the above objectives. Today, significant advances have been made in techniques of surface structuring at micro and nano scales, which enable significant modification of the surface wettability (cf. Figure 1).

In the present study, experiments were performed to determine the effects of surface wettability on the flow boiling heat transfer. The test channel is rectangular with hydraulic diameter of 0.96 mm and length of 180 mm. The confinement number is equal to 2.6, and hence the test channel is considered as microchannel according to Kew and Cornwell theory [1], but as minichannel according to Kandlikar and Grande [2] classification. The local wall temperature and heat flux are directly measured. The channel geometry was precisely determined in order to reduce measurement uncertainties. The sample surfaces are produced by well-controlled processes to ensure that from one to another surface, only the wettability is modified as the other parameters remain constant.

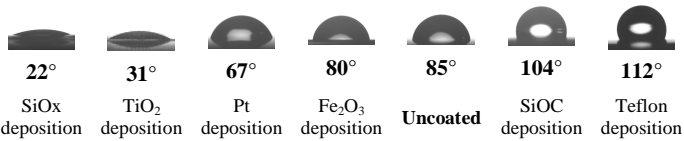

| 22° | 31° | 67° | 80° | 85° | 104° | 112° |
| SiOx deposition | TiO$_2$ deposition | Pt deposition | Fe$_2$O$_3$ deposition | Uncoated | SiOC deposition | Teflon deposition |

*Figure 1 : Static contact angles of 2-µL sessile water-droplets on stainless steel surfaces with and without nanoparticle deposition at 25°C [7].*

## 2. DESCRIPTION OF THE EXPERIMENTS

### a) Sample- surface fabrication

To determine the effects of surface wettability, it is necessary that all sample surfaces have the same geometry with only change in contact angle. In other words, they should have the same dimensions (length, width and thickness) and the same topography at microscale. The sample surfaces are thereby produced by deposition of nanoparticles through patterning masks using techniques of physical and chemical vapour depositions according to the following steps (cf. Figure 2).

*Step 1: deposition of titanium (Ti) layer.* The base substrate is a Pyrex wafer of 200 mm diameter and 1.1 mm thickness. This layer is used as a heating element. It consists of a rectangular track corresponding to the testing area and pads for electrical connections.

*Step 2: deposition of diamond-like carbon (DLC) layer.* This layer is used for electrical insulation.

*Step 3: deposition of nanoparticle layer.* This layer enables modification of the surface wettability in a larger threshold. The deposition materials are silicone oxide (SiOx) and carbon-doped silicon oxide (SiOC), respectively.

The testing area is 5 mm wide and 180 mm long (cf. Figure 3). It is heated by Joule effect from the metallic layer. Electrical wires are fixed on the electrical pads by mechanical support. Current and voltages of different sections of the testing area are measured by Agilent 34970A and a 0.01 Ω shunt, which has an accurately known resistance for determination of current by measurement of voltage.

- Pyrex wafer (200), thickness: 1.1 mm
- Titanium layer, thickness: 5 µm
- DLC electrical insulation layer, thickness: 0.5 µm
- Nanoparticle layer, thickness: < 30 nm

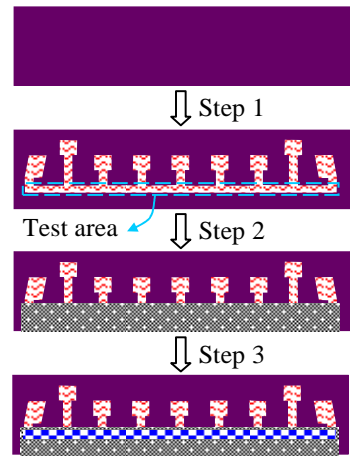

*Figure 2. Top view of fabrication procedure of sample surfaces.*

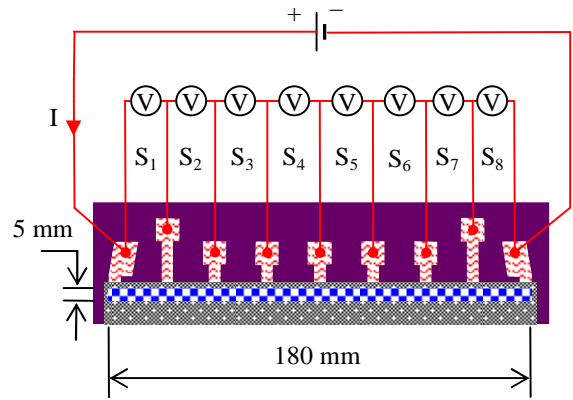

*Figure 3. Schematic view of electrical connexion.*

Once a sample surface is produced, stabilization of its deposition layers is made by annealing in a vacuum chamber at 300 °C for 3 hours. Afterwards, it is put inside a thermostat where the temperature was measured by a platinum probe of 0.1 °C accuracy. At steady state, the temperature of the sample surface could be determined from the temperature of the thermostat. For temperature between 20 °C and 90 °C, the electrical resistances of different sections of the sample surface are determined. In this way, the local wall temperature

 

at each section of the sample surface can be deduced from measurements of the electric resistance by using the Resistance/Temperature curve.

Applying the above method, four sample surfaces were fabricated and characterized such as:
- Titanium (Ti) surface made by step 1,
- Diamond-like carbon (DLC) surface made by step 1 and 2,
- Silicone oxide (SiOx) surface and carbon-doped silicon oxide (SiOC) surface made by step 1, 2 and 3.

*b) Experimental apparatus*

The channel is defined by putting a glass lid over the sample surface. This lid has a rectangular groove 5 mm wide, 0.5 mm high and is bonded to the Pyrex wafer by vacuum aspiration (cf. Figure 4). To prevent fluid leakage, silicone compound is placed around the external contact between the glass cap and the Pyrex wafer. The channel is thermally insulated with foam. Visualisation of the fluid flow can be made from the top of the glass lid.

The experimental setup is shown in Figure 5. It consists of a test section, a condenser with a cooling bath, a liquid pump (ISMATEC MCP_Z), a mass flowmeter (Micro Motion Elite MVD) and a pre-heater. A reservoir is used to store the fluid and to control the working pressure at atmospheric pressure.

The experimental facility is instrumented with an absolute pressure transducer (1 bar) to measure the pressure at the inlet of the test section, and a differential pressure transducer (100 mbar) to measure the pressure drop across the test section. The absolute pressures at the outlet of the condenser and at the inlet of the pre-heater are also measured. K-type thermocouples are inserted at different locations to measure the bulk fluid temperature.

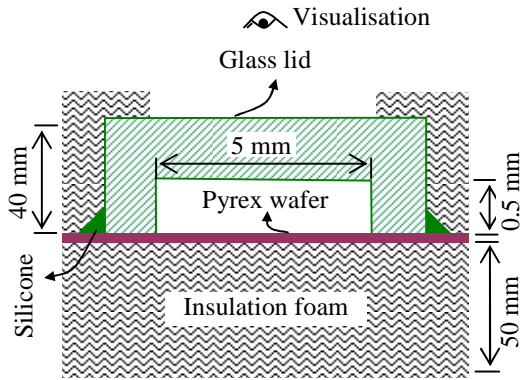

*Figure 4. Assembly of the test section.*

*c) Experimental procedure*

Before each test point, degassing of water is made by boiling at saturated temperature (100 °C) for two hours. Then, the desired flow rate is established and the electrical power is raised in steps lasting a few minutes each until a new steady state is achieved. The flow rate, current, voltages, pressures, and bulk temperatures are monitored and recorded at each power step. Flow visualisation was made by a high speed camera set at 500 or 1000 fps. The mass flux was set at 100 kg/m² s and the base heat flux was varied from 30 to 80 kW/m². All tests are performed under steady-state conditions.

The experimental data are recorded with a data logger (Agilent 34970A) connected to a computer.

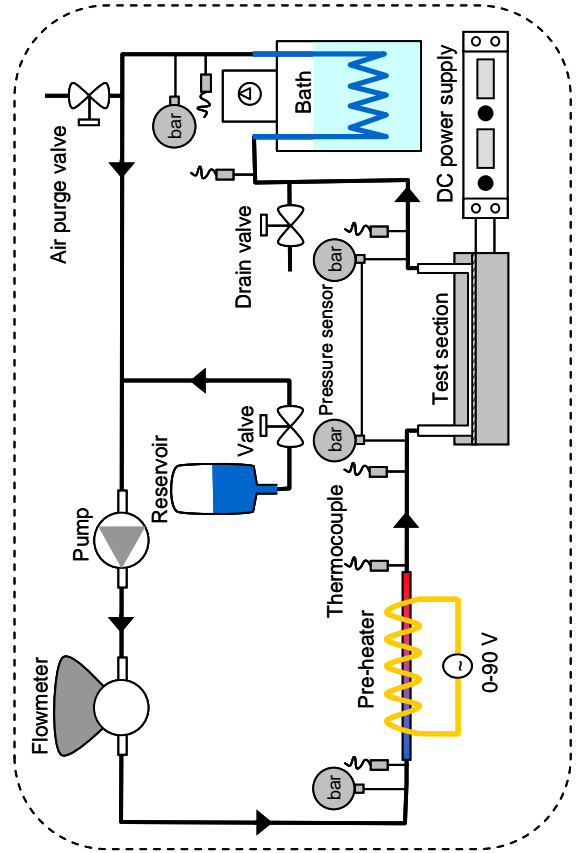

*Figure 5. Schematic view of the experimental apparatus.*

## 3. DATA REDUCTION

For a given test point, the local heat flux at section *i* of the sample surface is calculated as:

$$q_i = \frac{I V_i}{A_{h,i}} \quad (2)$$

where $I$ is the current, $V_i$ is the voltage and $A_{h,i}$ is the area of section *i*. The local heat transfer coefficient is determined from the local wall to bulk temperature difference and the heat flux as:

$$h_i = \frac{q_i}{T_{w,i} - T_{f,i}} \quad (3)$$

Under subcooled conditions, the bulk temperature is calculated from the inlet temperature and the heat added to the test section:

$$T_{f,i} = T_{f,in} + \frac{A_{h,i}(q_i - q_{loss,i})}{\dot{m} C_{p,i}} \quad (4)$$

where $q_{loss,i}$ is the heat loss at section *i*, which is mainly due to air convection near the test section and can be estimated by the energy balance in single-phase flow conditions.

 

The local saturation temperature, $T_{s,i}$, is determined by taking into account the drop in saturation temperature due to pressure drop, where the pressure drop is assumed to vary linearly along the test section. The variation of the vapour quality is calculated as:

$$\Delta x_i = \frac{A_{h,i}\,(q_i - q_{loss,i}) - \dot{m}\,C_{p,i}(T_{s,i} - T_{f,i})}{\dot{m}\,h_{lv,i}} \quad (5)$$

where $h_{lv}$ is the latent heat of vaporisation.

For average analysis, the average wall temperature is determined by the following relation:

$$\overline{T}_w = \sum_{i=1}^{N} A_{h,i} \times T_{w,i} \bigg/ \sum_{i=1}^{N} A_{h,i} \quad (6)$$

where $N$ is the number of sections of interest on the sample surface.

Thermodynamic properties of water are calculated with the computer code REFPROP 7.0, developed by NIST (2002). Experimental parameters and operating conditions are summarized in Table 1. The measurement uncertainties are estimated using the error propagation law suggested by Kline and McClintock [8].

*Table 1. Operating parameters and uncertainties*

| Parameter | Range | Uncertainty |
| --- | --- | --- |
| $D_h$ (mm) | 0.96 | ±0.02 mm |
| $G$ (kg/m²s) | 100 | ±2% |
| $P_{in}$ (mbar) | 1000 | ±0.1% |
| $P_{in} - P_{out}$ (mbar) | 0-100 | ±0.3% |
| $T_s$ (°C) | 100 | ±0.2 °C |
| $T_w$ (°C) | 100-120 | ±0.2 °C |
| $q$ (W) | 0-500 | ±2% |
| $h$ (kW/m² K) | 3-30 | ±4-7% |
| $X$ | -0.1-0.2 | ±2% |

## 4. SINGLE-PHASE FLOW VALIDATION TESTS

Validation tests are made by using the titanium surface (made by step 1 in Section 2a). To validate the test facility and the test section, the first step was to perform energy balance tests with highly-subcooled water flow (25 °C below the saturation temperature) at the inlet of the test section. The electrical heat loss from the sample surface to the fluid flow was less than 5% for base heat flux greater than 1 W/cm² and less than 3% for base heat flux greater than 2 W/cm². The electrical heat loss can be then neglected.

The second step was to perform single-phase liquid measurements of pressure loss and heat transfer to validate the measurement technique and the data reduction procedure. The fluid enters the test section at the ambient temperature (20 °C) at 100 kg/m² s. Figure 6 shows the evolution of the regular pressure loss with the Reynolds number. The experimental data show a relatively good agreement with the theoretical solutions of Shah and London [9] (maximum deviation of 7%). The evolution of the Nusselt number with the reduced flow length is shown in Figure 7. A good agreement between the experimental and the theoretical data is also obtained, with only a maximum deviation of 5%, even at low Reynolds numbers.

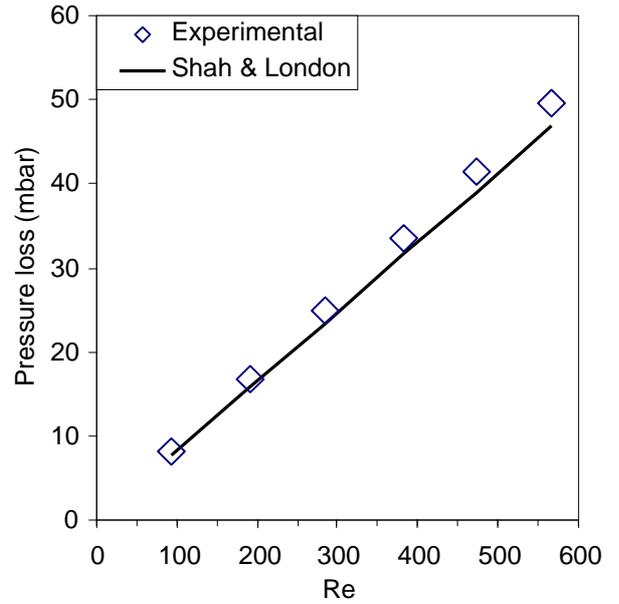

*Figure 6. Pressure loss vs. Reynolds number*

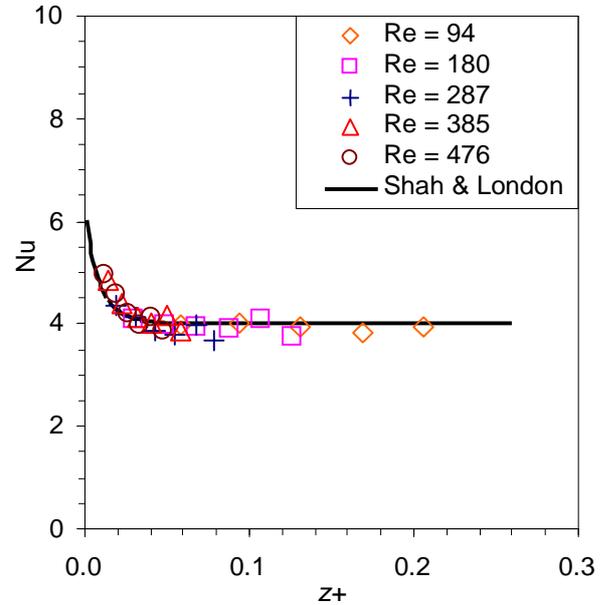

*Figure 7. Local Nusselt number vs. reduced length.*

## 5. FLOW BOILING EXPERIMENTAL RESULTS

The contact angles of water on the sample surfaces were measured by using the sessile drop technique with KRÜSS EasyDrop systems, as shown in Figure 8. Measurements were made in a cleanroom at room temperature.

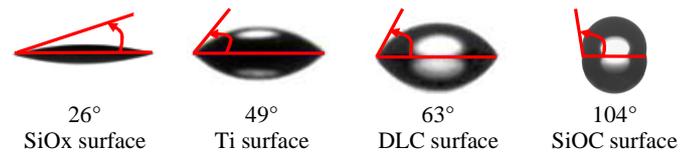

26°       49°        63°         104°
SiOx surface   Ti surface   DLC surface   SiOC surface

*Figure 8. Static contact angles of a water-droplet on the sample surfaces at room temperature.*

The SiOx surface shows a relatively high wettability; whereas the SiOC surface is an unwetted (hydrophobic) surface. The Ti and DLC surfaces are both wetted (hydrophilic) and have static contact angles ($\theta$) of 49° and 63°, respectively. The contact angle hysteresis $\Delta\theta$ of each sample surface is also determined by measurements of receding and advancing contact angles ($\theta_a$ and $\theta_r$, respectively). The results of contact angle measurements are summarized in Table 2.

*Table 2. Contact angle measurements*

| Surface | $\theta$ (°) | $\theta_a$ (°) | $\theta_r$ (°) | $\Delta\theta$ (°) |
|---|---|---|---|---|
| SiOx | 26 | 38 | 15 | 23 |
| Ti | 49 | 82 | 36 | 46 |
| DLC | 63 | 94 | 51 | 43 |
| SiOC | 104 | 108 | 96 | 12 |

Figure 9 shows the evolution of the heat transfer coefficient as a function of the vapour quality on different sample surfaces. In boiling condition, the heat transfer coefficient is in the range from 5000 to 30000 W/m² K. This is in agreement with observations of Sobierska et al. [10].

With a given heat flux, when boiling occurs, it is noticed that the heat transfer coefficient increases with increase in the vapour quality. For a vapour quality between 0.005 and 0.015, the heat transfer coefficient reaches its maximum value, and then decreases to a constant value. An increase in the heat flux leads to an increase in this constant value.

However, for SiOx surface which has a relatively high wettability, the heat transfer coefficient slightly changes when the vapour quality or the heat flux increase. This surface also shows the lowest heat transfer performance. For SiOC surface which is hydrophobic, nucleate boiling occurred even at negative vapour quality, i.e. at a fluid temperature lower than the saturation temperature.

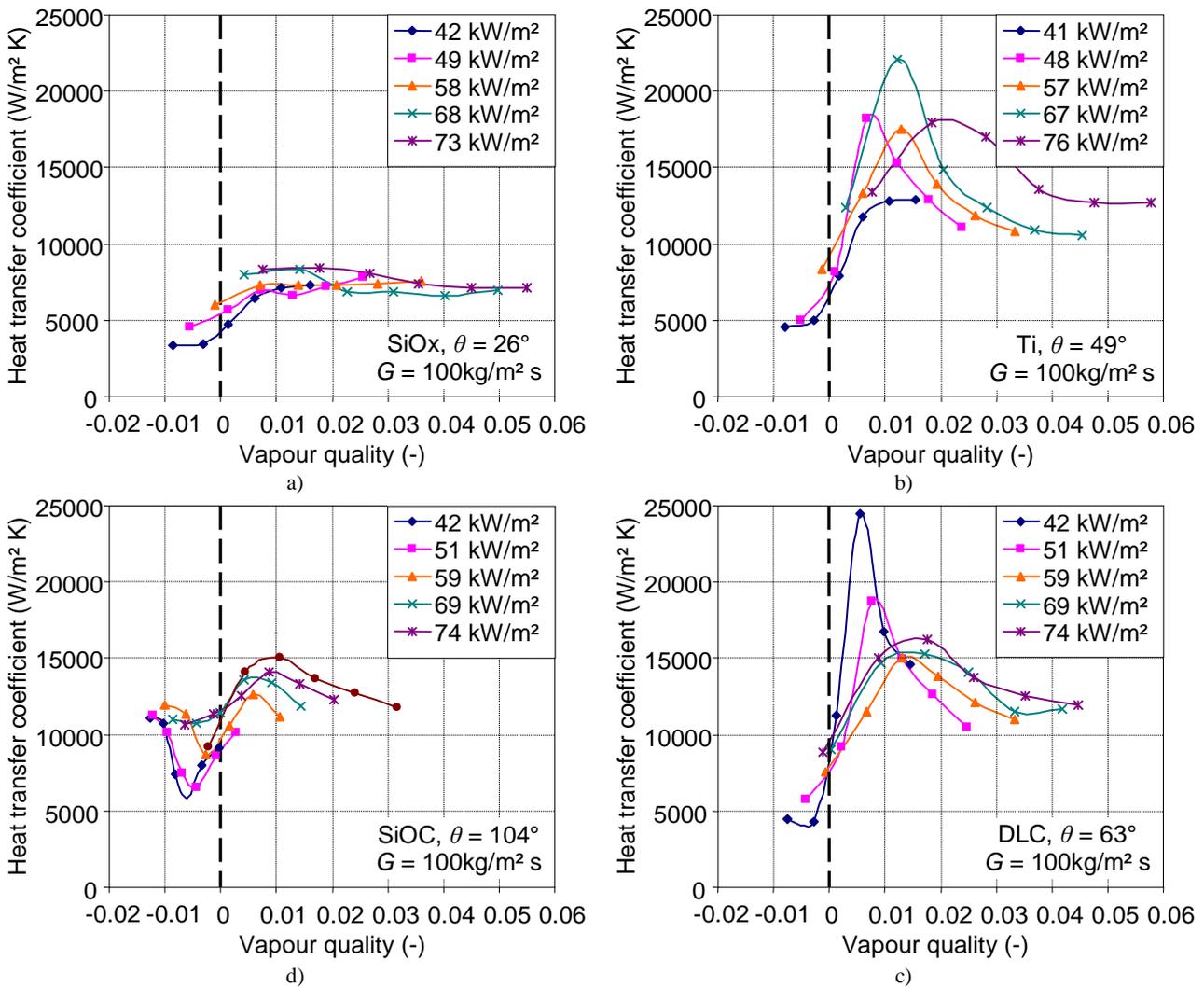

*Figure 9. Heat transfer coefficient vs. vapour quality on a) SiOx surface; b) Ti surface; c) DLC surface and d) SiOC surface.*

## 6. DISCUSSION

In order to clarify the boiling processes, images taken from the high speed camera were analysed. Figure 10 shows the flow patterns on the titanium surface. Three boiling flow patterns were identified: bubbly flow (BF), slug flow (SF) and semi-annular flow (SAF). The intermittent dryout was observed (by condensation of vapour on the top wall of the test channel) when the flow pattern changes from slug to semi-annular flow.

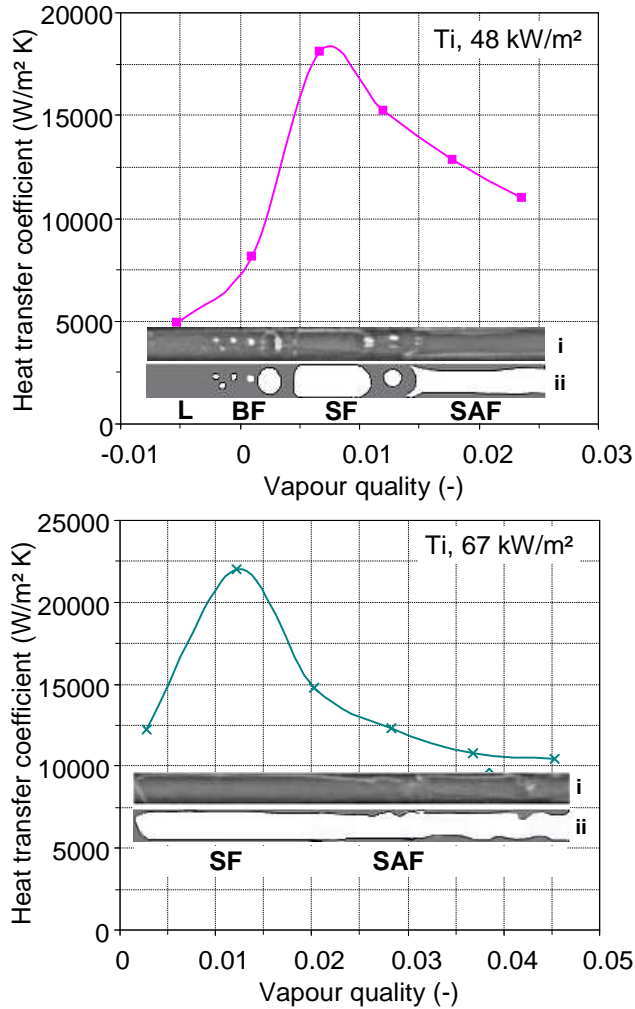

*Figure 10. Flow patterns on titanium surface with i) original image and ii) image with contour sharpening.*

The heat transfer coefficient significantly increases in the bubbly and slug flow configurations. Its maximum value would be obtained in slug flow when the liquid layer thickness reaches its minimum value. The heat transfer deterioration occurs when elongated bubbles in slug flow coalesce and form some churning liquid zones at the wall. This can be explained by the presence of the intermittent dryout observed in the semi-annular flow, as well as the increase of liquid film thickness in wavy conditions.

The experimental observations are also revealed by interpreting the boiling curves shown in Figure 11, where the wall temperature is taken as the average temperature defined by Eq. (6). It is shown that the superheat needed for onset of nucleate boiling (ONB) decreases when the contact angle increases.

For SiOC hydrophobic surface ($\theta = 103°$), nucleate boiling occurs even at a fluid temperature lower than the saturation one. This phenomenon could be explained by a large amount of gas trapped inside the sample-surface cavities. However, high rate of gas production leads to rapid bubble coalescence, causing deterioration of the heat transfer coefficient.

For SiOx highly-wetted surface ($\theta = 26°$), because of high superheat needed for the ONB, the heat transfer coefficient is relatively poor. Heat transfer would be mainly generated by conduction through the liquid layer, which is in contact with the heated surface. This layer is a result of balance of different factors such as the shear force at the liquid-vapour interface, the evaporation rate and the capillary effect. Since the surface is highly wetted, the capillary effect is thought to be the dominant factor. Thus, the thickness of the liquid layer is partially unchanged when the vapour quality or the heat flux increase. Therefore, the heat transfer coefficient remains almost constant as shown in Figure 9a.

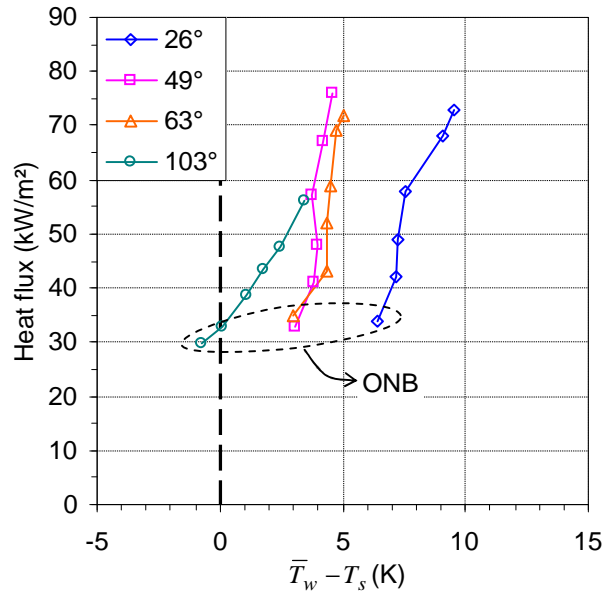

*Figure 11. Boiling curves for different surfaces having different contact angles.*

## CONCLUSION

The flow boiling heat transfer of water on nanostructured surfaces was determined. The experimental results show different flow patterns such as bubbly flow, slug flow, and semi-annular flow, which significantly impact the heat transfer. The general tendency of evolution of the heat transfer coefficient with the vapour quality is as follows: when boiling occurs, the heat transfer coefficient increases with the vapour quality. For a vapour quality between 0.005 and 0.015, the heat transfer coefficient reaches its maximum value, and then decreases before remaining constant.

It is also shown that the superheat needed for onset of nucleate boiling decreases when the contact angle increase. For SiOC hydrophobic surface, nucleate boiling occurs even at a negative vapour quality, i.e. at a fluid temperature lower than the saturation one. For SiOx highly wetted surface, the heat

 

transfer coefficient slightly changes when the vapour quality or the heat flux increase.